\documentclass[12pt]{article}
\usepackage{epsf}
\begin{document}
\def\beq{\begin{equation}}
\def\eeq{\end{equation}}
\def\bea{\begin{eqnarray}}
\def\eea{\end{eqnarray}}
\def\bbxsg{B\rightarrow X_s \gamma}
\def\as{\alpha_s}
\def\mz{M_z}
\def\ve{\vert}
\def\vel{\left|}
\def\ver{\right|}
\def\nnb{\nonumber}
\def\ga{\left(}
\def\dr{\right)}
\def\aga{\left\{}
\def\adr{\right\}}
\def\rar{\rightarrow}
\def\nnb{\nonumber}
\def\la{\langle}
\def\ra{\rangle}
\def\lla{\left<}
\def\rra{\right>}
\def\ba{\begin{array}}
\def\ea{\end{array}}
\def\ds{\displaystyle}
\def\lesssim{\mathrel{\mathpalette\vereq<}}
\def\vereq#1#2{\lower3pt\vbox{\baselineskip1.5pt \lineskip1.5pt
\ialign{$\m@th#1\hfill##\hfil$\crcr#2\crcr\sim\crcr}}}
\def\gtrsim{\mathrel{\mathpalette\vereq>}}
\def\alt{\lesssim}
\def\agt{\gtrsim}
\def\bos{\lower 0.5cm\hbox{{\vrule width 0pt height 1.2cm}}}
\def\boss{\lower 0.35cm\hbox{{\vrule width 0pt height 1.cm}}}
\def\aaa{\lower 0.cm\hbox{{\vrule width 0pt height .7cm}}}
\def\dol{\lower 0.4cm\hbox{{\vrule width 0pt height .5cm}}}
\newcommand{\dnp}{\delta^{{\scriptscriptstyle NP}}}
\newcommand{\bsg}{\mbox{$b \to s \gamma\,$}}
\newcommand{\bxsg}{\mbox{$b \to X_s \gamma\,$}}
\newcommand{\bsGAM}{\mbox{$B \to X_s \gamma\,$}}
\newcommand{\br}{\mbox{${\rm BR}(b \to s \gamma)\,$}}
\newcommand{\brx}{\mbox{${\rm BR}(b \to X_s \gamma)\,$}}
\newcommand{\bbrx}{\mbox{${\rm BR}(\overline{B}\to X_s\gamma)\,$}}
\newcommand{\bSL}{\mbox{$b \to c l \nu_l$}}
\newcommand{\brSL}{\mbox{${\rm BR}(b \to c l \nu_l)\,$}}
\newcommand{\thb}{\mbox{$t\to H^+ b\,$}}
\newcommand{\mh}{m_\smallh}
\newcommand{\smallSL}{{\scriptscriptstyle SL}}
\newcommand{\barbbrx}{\mbox{${\rm BR}(B\to\overline{X_s}\gamma)\,$}}
\newcommand{\muw}{\mu_\smallw}
\newcommand{\mut}{\mu_t}
\newcommand{\mub}{\mu_b}
\newcommand{\hd}{{\overline D}}
\newcommand{\imx}{\rm{Im\,X}}
\newcommand{\rex}{\rm{Re\,X}}
\newcommand{\xx}{\mbox{$\rm{X}$}}
\newcommand{\yy}{\mbox{$\rm{Y}$}}
\newcommand{\zz}{\mbox{$\rm{Z}$}}
\newcommand{\xys}{{\rm{X\,Y}^\ast}}
\newcommand{\rxys}{{\rm Re}\,(\xx\,\yy^\ast)}
\newcommand{\ixys}{{\rm Im}\,(\xx\,\yy^\ast)}
\title{\Large {\bf Constraining Fourth Generation with $B\rightarrow X_s \gamma$}}
\author{\vspace{1cm}\\{\small L. Solmaz \thanks{e-mail:lsolmaz@photon.physics.metu.edu.tr.}}\\
{\small Physics Department, Middle East Technical University} \\
{\small 06531 Ankara, Turkey} }
\date{}
\begin{titlepage}
\maketitle
\thispagestyle{empty}
\begin{abstract}
  Using the  theoretical and experimental results on $\bsGAM$, a
four-generation SM is analyzed to constrain the combination of
the $4\times 4$ Cabibbo-Kobayashi-Maskawa factor $V_{t^\prime
s}^\ast V_{t^\prime b}$ as a function of the $t^\prime$--quark
mass. It is observed that the results for the above--mentioned
physical quantities are essentially different from the previous
predictions  for certain solutions of the CKM factor. Influences
of the new  model is used to predict CP violation in $\bsGAM$
decay at the order of $A_{CP}=5~\% $, stemming from the
appearance of complex phases of  $V_{t^\prime
s}^\ast V_{t^\prime b}$ and of Wilson coefficients $C_7$,
$C_8$, in the related process. The above mentioned physical
quantities can serve as efficient tools in search of the fourth
generation.
\end{abstract}
\end{titlepage}

\section{Introduction}
~~Today, despite the success of the Standard Model (SM), from the
theoretical point of view, it is incomplete. Number of
generations of fermions can be mentioned as one of the open
problems of SM, for which we do not have a clear argument to
restrict the SM to three known generations. Mass of the extra
generations, if ever  exists, can be extracted from the
measurements of  neutrino  experiments, which set a lower bound
for  extra generations ($m_{\nu_4}> 45~ GeV$) \cite{Ref1}.

The idea of generalizing SM is not a new one. Probable effects of
extra generations  was studied in many works
\cite{Ref2}--\cite{Ref16}. Generalizations of the SM can be used
to introduce  a new family, which was performed  previously \cite{Ref17}.
Using similar techniques, one can  search fourth generation
effects in  B  meson decays. The existing electroweak data on the
$Z$--boson parameters, the $W$ boson and the top quark masses
excluded the existence of the new generations with all fermions
heavier than the $Z$ boson mass \cite{Ref16}, nevertheless, the
same data allows few extra generations, if neutral
leptons  have masses close to $50~GeV$.

$\bsGAM$  is one of the most promising areas in search of the
fourth generation, via its indirect loop effects, which was
performed previously \cite{Ref7,Ref8}. This decay is one of the
most appropriate candidates to be searched in the extensions of
SM, since we have solid experimental and theoretical background
for the process under consideration.

In this work we study the contribution of the fourth generation
in the rare $\bsGAM$ decay, to obtain constrains on the parameter
space of the fourth generation. Our basic assumption is to fill
the gap between theoretical and experimental results of $\bsGAM$,
with the fourth generation. Of course, due to the mentioned
assumption, decay width will change at the order of difference
between theoretical and experimental results, however, predicted
CP asymmetry is interesting when  SM contribution is neglected.
As it is well known, new physics effects can manifest themselves
through the Wilson coefficients and their values can be different
from the ones in the SM \cite{Ref18,Ref19}, as well as through
the new operators \cite{Ref20}. Note that the inclusive $B \rar
X_s \gamma$ decay have already been studied with the inclusion of
the fourth generation  \cite{Ref21,Ref210} to constrain
$V_{t^\prime b}^\ast V_{t^\prime s}$. The restrictions of the
parameter space of nonstandard models based on LO analysis  are
not as sensitive as in the case of NLO analysis. Therefore we
preferred to work at NLO, for the decay under consideration.

On the experimental side, values related with the $\bsGAM$ are
well known. First measurement of the $\bsGAM$ was performed by
CLEO collaboration, leading to CLEO branching ratio
~\cite{cleoincl} \beq \bsGAM = (2.32\pm 0.57\pm 0.35)\times
10^{-4}\,. \label{CLEOres} \eeq In 1999, CLEO has presented an
improved result \cite{cleonew}\beq \bsGAM = (3.15\pm 0.35\pm
0.32\pm 0.26)\times 10^{-4}\,. \label{CLEOres2} \eeq The errors
are statistical, systematic, and model dependent respectively. The
rate measured by ALEPH \cite{aleph} is consistent with the CLEO
measurement. There exists also  results of   BELLE  with a larger
central value \cite{belleexpo}: \beq \bsGAM =(3.37\pm 0.53\pm
0.42\pm 0.54 )\times 10^{-4} \,. \label{BELLEres} \eeq Observing
CP asymmetry in the decay  $B \rightarrow  X_s \gamma$ is
interesting, presented by CLEO collaboration recently
\cite{cleoCP}
\begin{equation}
A_{CP}(B \rightarrow  X_s \gamma) = (-0.079 \pm 0.108 \pm 0.022)
\times (1.0 \pm 0.030) \, ,
\end{equation}
for which prediction of the SM is $0.6 ~\%$ \cite{tobiash}.

On the theoretical side, situation within  and beyond the SM is
well settled. A collective theoretical effort  has led to the
practical determination of $\bsGAM$ at the NLO,  which  was
completed recently, as a joint effort of many different groups
(\cite{AG91},\cite{Pott}, \cite{GHW}, \cite{Adel},
\cite{Mikolaj},\cite{GH}). For a recent review, to complete the
computation of NLO  QCD corrections, we refer to ref.
\cite{completed} and references therein. It is necessary to have
precise calculations also in the extensions of the SM, which was
performed for certain models \cite{extensions}. With the
appearance  of more accurate data we will be able to provide
stringent constraints on the free parameters of the models beyond
SM. We can state that, the aim of the present paper is to obtain
such  constraints when the fourth generation is considered.

 The paper is
organized as follows. In section 2, we present the necessary
theoretical expressions for the $\bsGAM$  decay in the SM with
four generations, where we investigated the effect of introducing fourth generation
at different scales upon branching ratio and CP
asymmetry. Section 3 is devoted to the numerical analysis and our
conclusion.

\section{Theoretical results}
We use the framework of an  effective low-energy theory, obtained
by  integrating  out heavy degrees of freedoms, which in our case
W-boson and top quark and an additional $t^{\prime}$. Mass of the
$t^{\prime}$ is at the order of $m_W$. In this approximation the
effective Hamiltonian relevant for $b\rightarrow s \gamma$ decay
reads \cite{misiak1,JH93}

\bea {\cal H}_{eff} &=& \frac{4 G_F}{\sqrt{2}}V_{ts}^\ast  V_{tb}
\sum_{i=1}^{8} { C}_i(\mu) \, { O}_i(\mu)~, \eea

where $G_F$ is the Fermi coupling constant  $V$ is the
Cabibbo-Kobayashi-Maskawa (CKM) quark mixing matrix, the the full
set of the operators ${O}_i(\mu)$ and the corresponding
expressions for the Wilson coefficients ${C}_i(\mu)$ in the SM
can be found in (\cite{AG91}--\cite{Mikolaj}).

 In the model under consideration, the fourth generation is
introduced in a similar  way the three generations are introduced
in the SM, no new operators appear and clearly the full operator
set is exactly the same as in SM \cite{JH93}. The fourth
generation changes  values of the Wilson coefficients
$C_7(\mu_W),~C_8(\mu_W)$, via virtual exchange of the fourth
generation up quark $t^\prime$ at matching  scale. With the
definition $\lambda_{i}=V_{i s}^{\ast} V_{i
b},~i=\{u,~c,~t,~t^\prime\}$,
  the above mentioned Wilson coefficients, can be written in the following
form

\bea
 C_7^{eff}(\mu_W) &=& C_7^{SM}(\mu_W) + \frac{\lambda_{t^\prime}}
{\lambda_{t}} C_7^{New} (\mu_W) ~, \nnb \\
C_8^{eff}(\mu_W) &=& C_8^{SM}(\mu_W) + \frac{\lambda_{t^\prime}} {\lambda_{t}} C_8^{New} (\mu_W)
~,   \eea

 where the last terms in these expressions describe the
contributions of the $t^\prime$ quark to the Wilson coefficients
and $V_{t^\prime s}^{\ast}$ and $V_{t^\prime b}$ are the two elements of
the $4\times 4$ Cabibbo--Kobayashi--Maskawa (CKM) matrix.
 The explicit forms of the
$C_i^{New}$ can easily be obtained from the corresponding Wilson
coefficient expressions in SM by simply substituting $m_t \rar
m_{t^\prime}$ (see \cite{Ref23,Ref25}). Neglecting the $s$ quark
mass we can  define the Wilson coefficients at the matching scale,
where the LO functions are :
 \bea
 C_{7}^{SM}  & = & \frac{x}{24} \,
 \frac{-8x^3+3x^2+12x-7+(18x^2-12x) \ln x}{(x-1)^4}  \, \nonumber     \\
 C_{8}^{SM}  & = & \frac{x}{8} \,
 \frac{-x^3+6x^2-3x-2-6x \ln x}{(x-1)^4}  ,\,
\label{wclosm}
\eea
where ($x=m_t^2/m_W^2$).

In the  calculations we used the NLO theoretical expressions, and different
experimental values to constraint the $\lambda_{t^\prime}$ paramater.
Since extended models are very sensitive to NLO corrections, we used the NLO expression for the branching ratio
of the
radiative decay $B\to X_s \gamma$, which has been presented in
ref.~\cite{misiak1}: \bea  BR(B\to X_s \gamma )&=&BR(B \to X_c e
\bar{\nu}_e)\left| \frac{V_{ts}^* V_{tb}}{V_{cb}}\right|^2
\frac{6 \alpha_e}{\pi f(z)\kappa(z)}\frac{\bar
m_b^2(\mu_b)}{m_b^2}
\nonumber \\
&& \times \left( |D|^2 + A \right)  \left(
1-\frac{\delta^{NP}_{SL}}{m_b^2}
+\frac{\delta^{NP}_{\gamma}}{m_b^2}+\frac{\delta^{NP}_{c}}{m_c^2}\right)
~. \label{br} \eea

Explicit forms of virtual, bremsstrahlung  and
non-perturbative  parts of Eq. (\ref{br}) can be found in
\cite{misiak1, completed} and references therein. In the numerical analysis we
 obtained $B{\to}X_s\gamma$ branching ratio in the
Standard Model  $BR(\bsGAM)=(3.48 \pm 0.33)\times 10^{-4}$, which remains in agreement with the previous literature. But
we considered only the central value in our analysis, with the expectation of absorbing errors into the different experimental values.

To obtain quantitative results we need the value of the fourth
generation CKM matrix element $\lambda_{t^{\prime}}$. For this
aim following \cite{Ref21}, we will use the experimental results
of the decays $BR (\bsGAM)$ and  determine the fourth generation
CKM factor $\lambda_{t^{\prime}}$ When we consider the possible
effects of the fourth generation, we demanded the theoretical
value to be equal to the experimental values presented in the
previous section. Which can be expressed as \bea
 BR(B\rightarrow X_s \gamma)_{4th}&=&\{2.66,~3.15,~3.37\}.
\eea
 Theoretical results of the  branching ratio for
  $m_{t^{\prime}}={75,...,500} ~GeV$ values are obtained as as function of $\lambda_{t^\prime}$.
   Notice that  in the  expressions related with $BR(B\rightarrow X_s \gamma)_{4th}$, theoretical and experimental
   results are  multiplied by a factor of  $10^4$ . For instance
   when we chose $m_{t^{\prime}}=75$ GeV, and use the approach of
   Eq. (6):
\bea  BR(B\rightarrow X_s \gamma)_{4th}&=&
0.654502 \,+\,6.69962\,\lambda_{t^{\prime}}+\,20.3501\,\lambda^2_{t^{'}} +\, \nonumber\\
&& 0.396254\,|-0.305738-\,1.87828\,\lambda_{t^{\prime}}|^2 +\, \\
&& 23.9926
\,|(\,-0.340878-\,0.0154077\,i)-(\,1.64283+\,0.05443\,i){\lambda_{t^{\prime}})|^2}
\,.\nonumber \eea
  When   $\lambda_{t^{\prime}}$ is neglected
 branching ratio reduces to the re-scaled central value 3.48 of SM prediction.  During the calculations we obtained similar
  expressions for different $m_{t^\prime}$ values.
 It suffices to present the case of a very heavy quark,  for $m_{t^{\prime}}=~500~GeV$ :
\bea  BR(B\rightarrow X_s \gamma)_{4 th}&=&
0.654502 \,+\,20.9868\,\lambda_{t^{\prime}}+\,198.863\,{\lambda_{t^{\prime}}}^2 +\, \nonumber\\
&& 0.396254\,|-0.305738-\,5.95621 \,\lambda_{t^{\prime}}|^2 +\, \\
&& 23.9926
\,|(\,-0.340878-\,0.0154077\,i)-(\,5.1664+\,0.11899\,i){\lambda_{t^{\prime}})|^2}
\,.\nonumber \eea
In the numerical analysis, as a first step,
${\lambda_{t^{\prime}}}$ is assumed real and constraints are
obtained as a function of mass of the extra generation top-quark
$m_{t^{\prime}}$, and the values are presented in tab. (1) and can be obtained from fig.(1).
Those values can also  be extracted from the  figures $(3,4,~and~ 5)$ (a) where the solution is the intersection
point on the  $BR_{\gamma}=1$ line . Notice
that in the figures we normalized branching ratio to 1, using the
experimental values 2.66, 3.15 and 3.37 respectively, hence $\lambda_{t^\prime}$ values can be obtained from the
 intersection points and this is true for all $figures$ except the ones related with $A_{CP}$ , presented in the following  subsections.

\begin{figure}
\vskip .5 cm
    \includegraphics{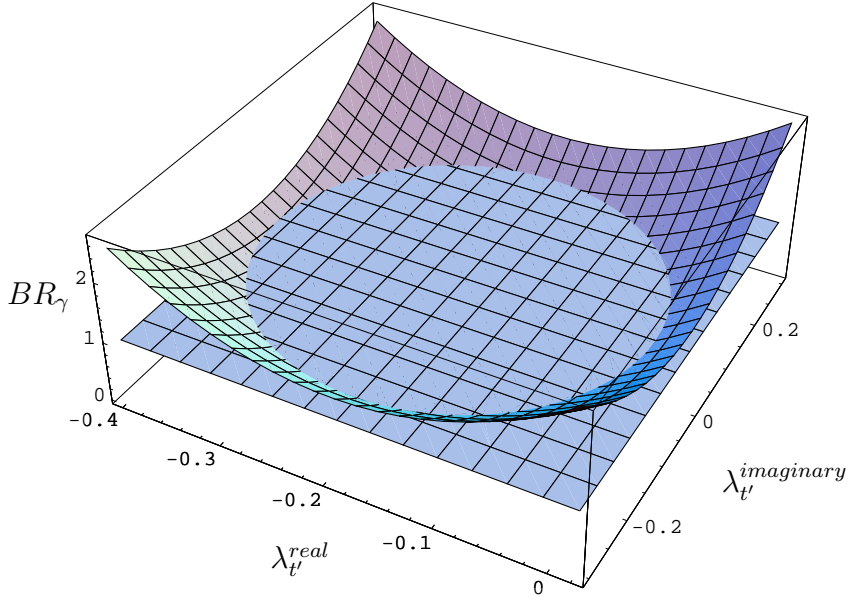}
    \vskip 1.5 cm \hskip 1 cm
\title{$BR_{\gamma}$}
\vskip 2 cm \hskip 10.5 cm
\title{$\lambda_{t^{\prime}}^{imaginary}$}
\vskip .5 cm \hskip 4.5 cm
\title{$\lambda_{t^{\prime}}^{real}$}
   \vskip .1 cm \caption{$BR(B\rightarrow X_s\gamma)$ normalized to 1 with the experimental value $BR(\bsGAM)=3.15$, in order to extract
   possible   values of  ${\lambda_{t^{\prime}}}$, for $m_{t^\prime}=75$ GeV.
   Constraints are obtained for Eq. (6), and can be inferred from the emerging circle.}
\end{figure}

\begin{figure}
\vskip .5 cm
    \includegraphics{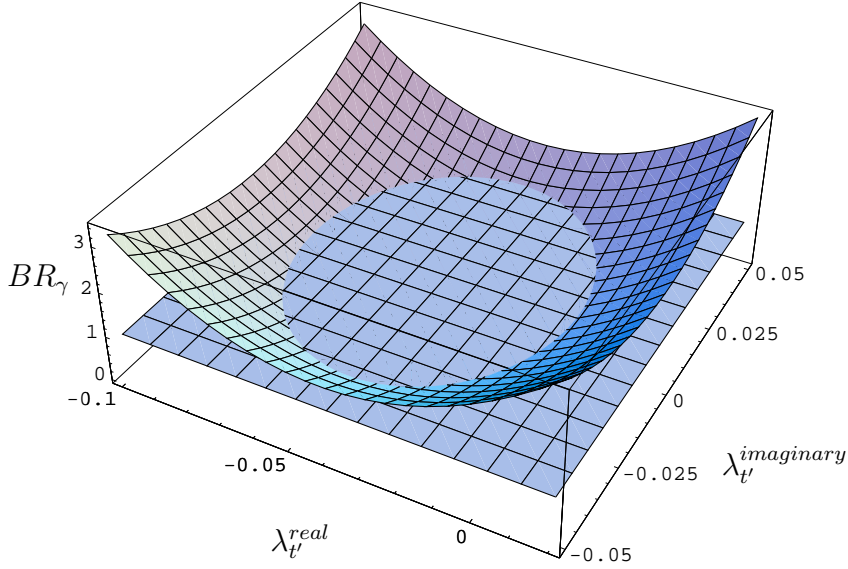}
    \vskip 1.5 cm \hskip 1 cm
\title{$BR_{\gamma}$}
\vskip 2 cm \hskip 10.5 cm
\title{$\lambda_{t^{\prime}}^{imaginary}$}
\vskip .5 cm \hskip 4.5 cm
\title{$\lambda_{t^{\prime}}^{real}$}
   \vskip .1 cm \caption{As in Fig.1. $BR(B\rightarrow X_s\gamma)$ normalized to 1 with the experimental value $BR(\bsGAM)=3.15$,
    but, constraints are obtained for Eq. (12)}
\end{figure}
\begin{figure}
\vskip .5 cm
    \includegraphics{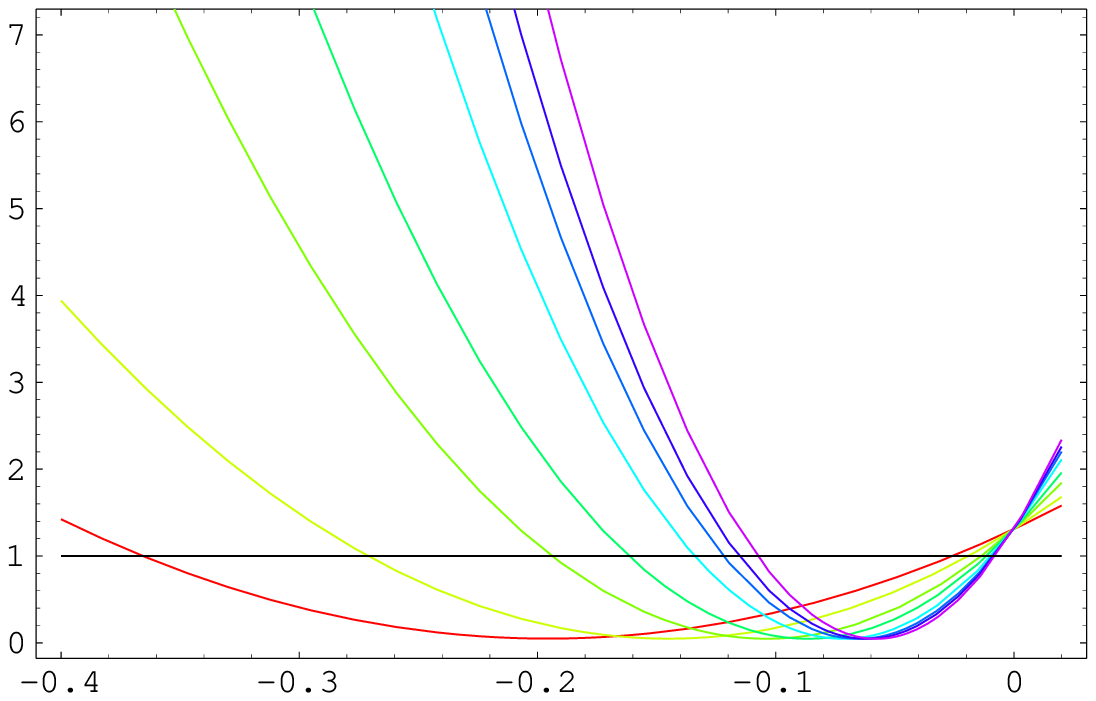}
    \hskip 7.5 cm
    \includegraphics{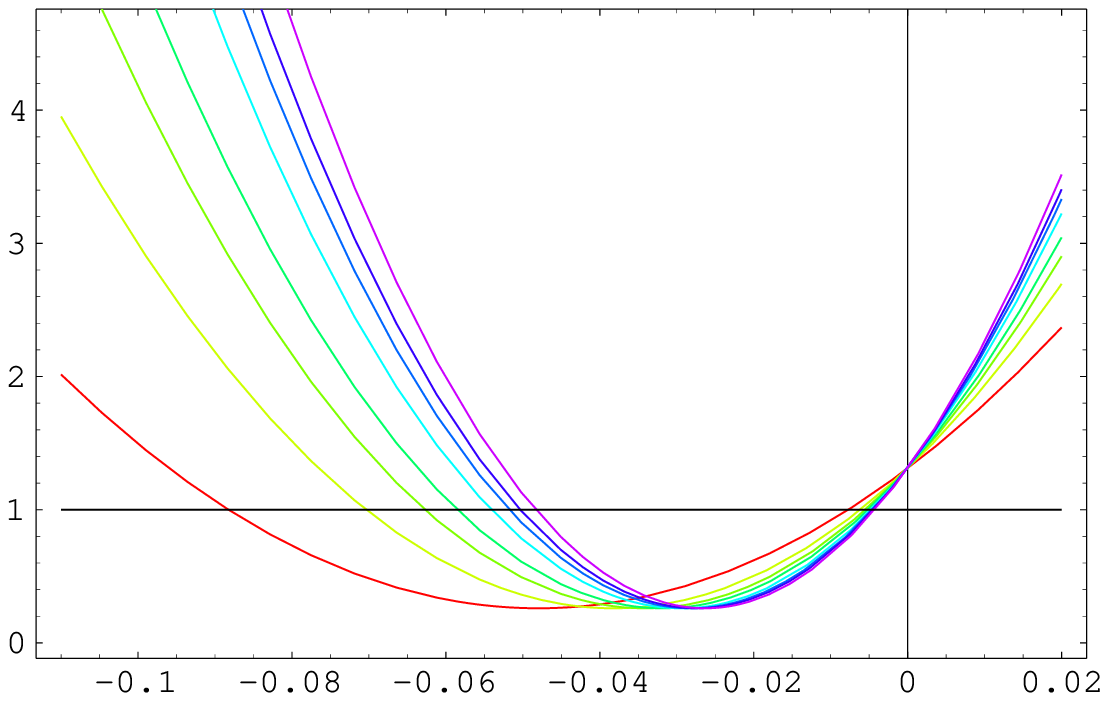}
    \vskip 2 cm \hskip .1 cm
\title{$BR_{\gamma}$}
\vskip 3 cm \hskip 5 cm
\title{(a)~~~~~~~~~~~~~~~~~~~~~~~~~~~~~~~~~~~~~~~~~~~~~~~~~~(b)}
   \vskip .1 cm \caption{$BR(B\rightarrow X_s\gamma)_{4th}$ normalized to 1, with the experimental value $BR(\bsGAM)=2.66$.
   Red line stands for $m_{t^\prime}=75~GeV$, pink one denotes $m_{t^\prime}=500~GeV$, other masses are in this range respectively.
   Notice that in the figures ${\lambda_{t^{\prime}}}$ values are assumed real. Fig. (a). is related with  Eq. (6) likewise,
    Fig. (b). is related with  Eq. (12). }
\end{figure}

\begin{figure}
\vskip .5 cm
    \includegraphics{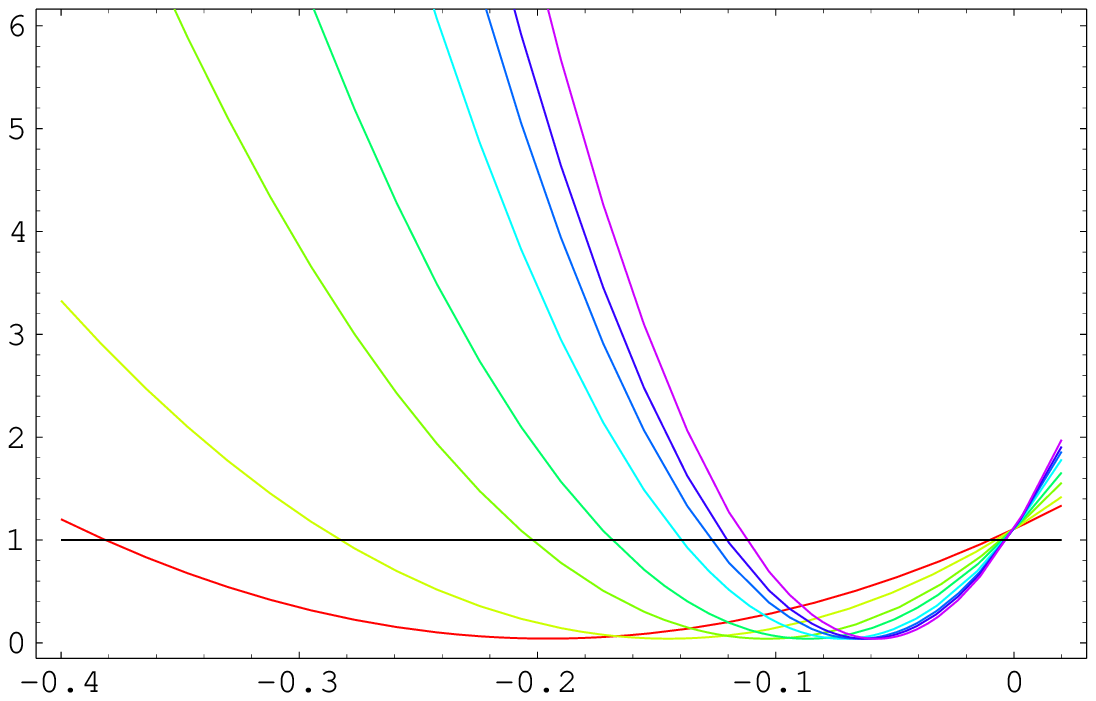}
    \hskip 7.5 cm
    \includegraphics{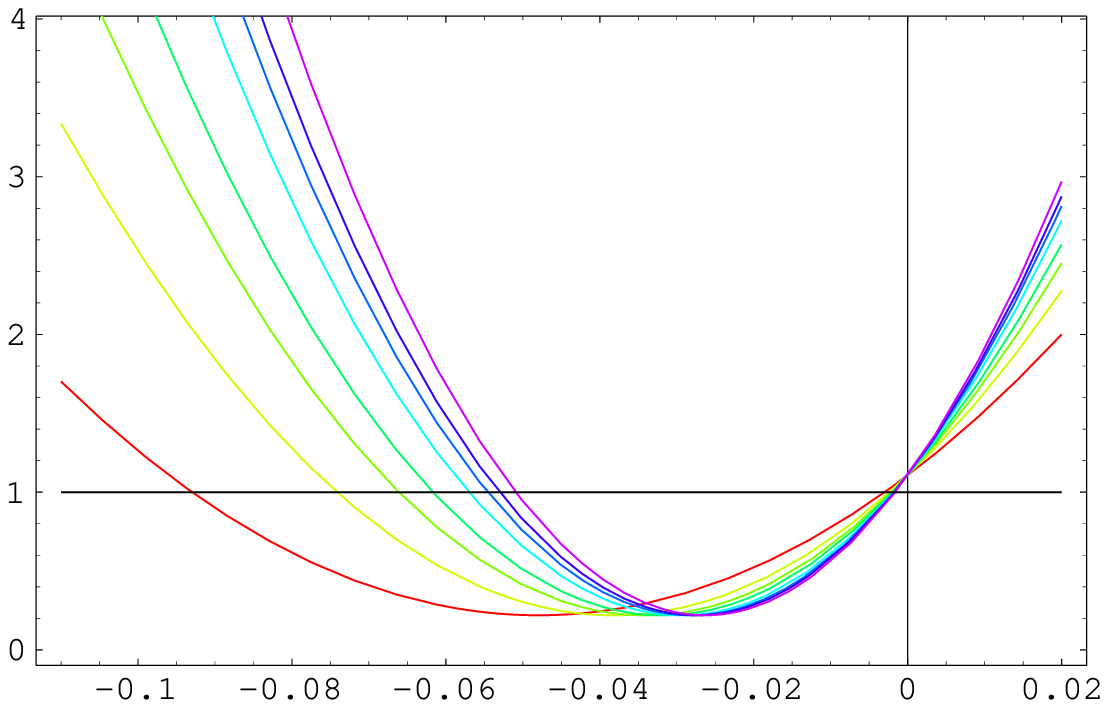}
    \vskip 2 cm \hskip .1 cm
\title{$BR_{\gamma}$}
\vskip 3 cm \hskip 5 cm
\title{(a)~~~~~~~~~~~~~~~~~~~~~~~~~~~~~~~~~~~~~~~~~~~~~~~~~~(b)}
   \vskip .1 cm \caption{The same as Fig.3., for the experimental value $BR(\bsGAM)=3.15$}
\end{figure}

\begin{figure}
\vskip .5 cm
    \includegraphics{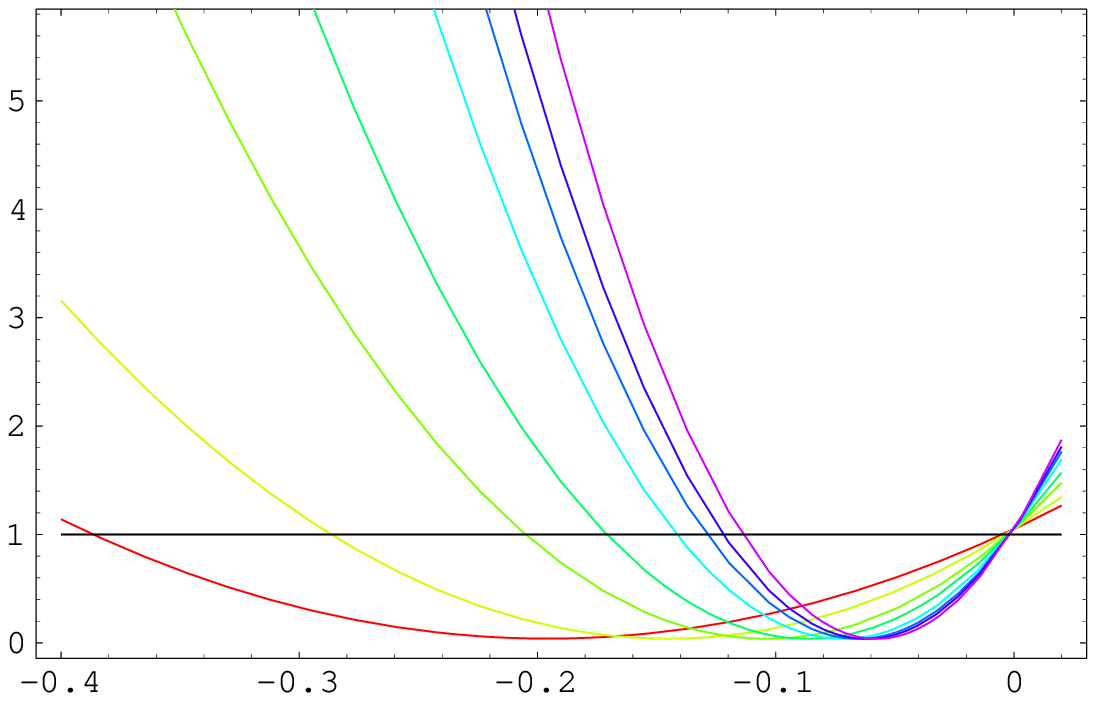}
    \hskip 7.5 cm
    \includegraphics{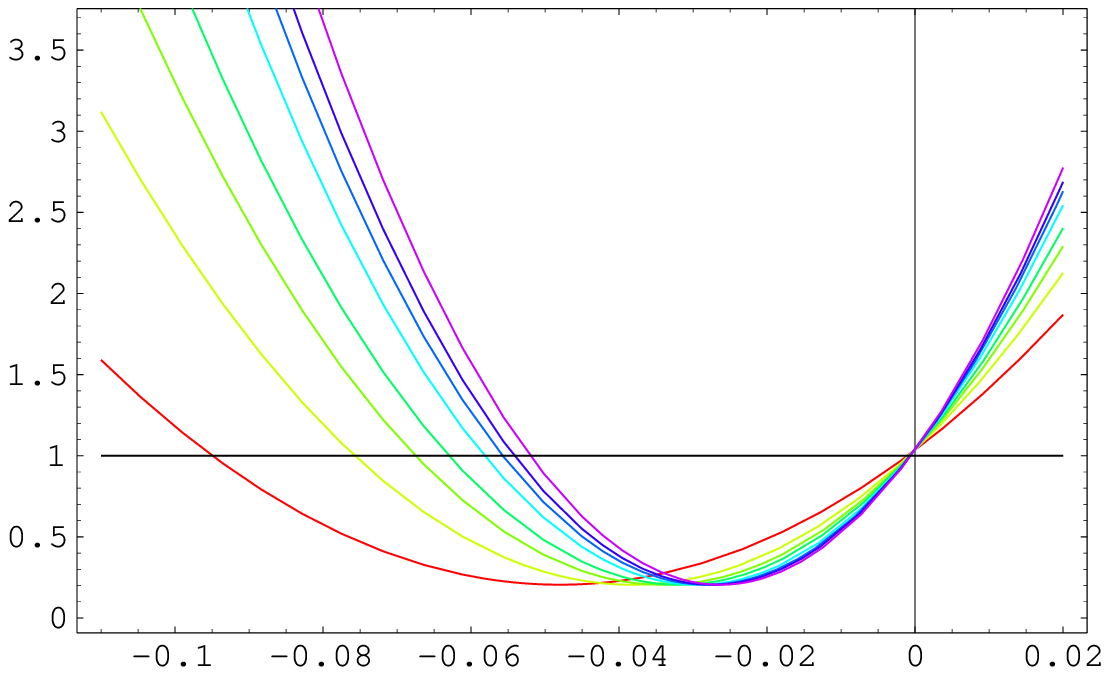}
    \vskip 2 cm \hskip .1 cm
\title{$BR_{\gamma}$}
\vskip 3 cm \hskip 5 cm
\title{(a)~~~~~~~~~~~~~~~~~~~~~~~~~~~~~~~~~~~~~~~~~~~~~~~~~~(b)}
   \vskip .1 cm \caption{The same as Fig.3., for the experimental value $BR(\bsGAM)=3.37$}
\end{figure}

We also performed a  very similar analysis for introducing the fourth generation effects at the $\mu_{b}$ scale to see the
difference between the previous results. Following \cite{Ref21} it can be written as follows:

\bea
 C_7^{eff}(\mu_b) &=& C_7^{SM}(\mu_b) + \frac{\lambda_{t^\prime} }
{\lambda_{t}}
 C_7^{New} (\mu_b) ~, \nnb \\
C_8^{eff}(\mu_b) &=& C_8^{SM}(\mu_b) + \frac{\lambda_{t^\prime}} {\lambda_{t}} C_8^{New} (\mu_b)
~,   \eea
Using Eq.(9), and demanding theretical results to be equal to the experimental  results again,
 we obtained following expression for  $m_{t^{\prime}}=75$:
\bea
  BR(B\rightarrow X_s \gamma)_{4th}&=& \,0.691542 +\, 23.9926\, \times \nonumber \, \\
  &&|(-0.340878-0.0154077 \,i) -(8.13033+0.423786 \,i)\lambda_{t^{\prime}}|^2
 \,\eea
As another example for $m_{t^{\prime}}=500$ we obtained
\bea
  BR(B\rightarrow X_s \gamma)_{4th}&=& \,0.691542 +\, 23.9926\, \times \nonumber \, \\
  &&|(-0.340878-0.0154077 \,i) -(12.4566\,+0.484564 \,i)\lambda_{t^{\prime}}|^2
 \,.\eea

 It is interesting to notice that, if we assume
 $\lambda_{t^{\prime}}$ can have imaginary parts, experimental values can  also be satisfied.
  This case is presented with a graphical solution
 in figure (2) for $m_t=75$ and the decomposition
$\lambda_{t^{\prime}}=\lambda_{t}^{real}\,+\,i\,\lambda_{t}^{imaginary}$. When $\lambda_{t^\prime}$ is assumed real
constraints can be extracted from figures (3,4,~and ~5) (b) on the normalized line. Real and imaginary parts or this approach
is presented in tables (2) and (3) respectively.

\begin{table}[ht]
\renewcommand{\arraystretch}{2}
\addtolength{\arraycolsep}{5pt}
$$
\begin{array}{|c|c|c|c|c|c|c|c|}\hline
\multicolumn{8}{|c|}{BR(\bsGAM)=2.66\times 10^{-4}}\\
\hline {m_{t^\prime}~(GeV)} & 75 & 100 & 150 & 200 & 300 & 400 & 500
\\
\hline \hline \lambda_{t^{\prime}}^{(I)}\times 10^{-1}
 &-3.63&-2.85&-2.04&-1.72&-1.42&-1.29&-1.22 \\ \hline
\lambda_{t^{\prime}}^{(II)}\times 10^{-3}
&-1.01&-0.75&-0.54&-0.45&-0.37&-0.34&0.32

\\ \hline
\multicolumn{8}{|c|}{BR(\bsGAM)=3.15\times 10^{-4}}\\
\hline {m_{t^\prime}~(GeV)} & 75 & 100 & 150 & 200 & 300 & 400 & 500
\\
\hline \hline \lambda_{t^{\prime}}^{(III)}\times 10^{-1}
&-3.90&-2.90&-2.08&-1.74&-1.45&-1.31&-1.25 \\ \hline
\lambda_{t^{\prime}}^{(IV)}\times 10^{-3}
&-3.4&-2.5&-1.8&-1.5&-1.2&-1.1&-1.1
\\ \hline
\multicolumn{8}{|c|}{BR(\bsGAM)=3.37\times 10^{-4}}\\
\hline {m_{t^\prime}~(GeV)} & 75 & 100 & 150 & 200 & 300 & 400 & 500
\\
\hline \hline \lambda_{t^{\prime}}^{(V)}\times 10^{-1}
&-3.67&-2.73&-1.96&-1.63&-1.35&-1.23&-1.12 \\ \hline
\lambda_{t^{\prime}}^{(VI)}\times 10^{-3}
&-2.6&-1.9&-1.4&-1.1&-1.0&-0.9&-0.8
\\ \hline\hline
\end{array}
$$
\caption{The numerical (real parts only) values of $\lambda_{t^{\prime}}$ for different values of the $m_{t^\prime}$ --quark mass and
experimental values . The
superscripts $(I),...,(VI)$ correspond to first and last
solutions of Eq. (9) with the approximation of Eq. (12). }
\renewcommand{\arraystretch}{1}
\addtolength{\arraycolsep}{-5pt}
\end{table}

\begin{table}[ht]
\renewcommand{\arraystretch}{2}
\addtolength{\arraycolsep}{5pt}
$$
\begin{array}{|c|c|c|c|c|c|c|c|}\hline
\multicolumn{8}{|c|}{BR(\bsGAM)=2.66\times 10^{-4}}\\
\hline {m_{t^\prime}~(GeV)} & 75 & 100 & 150 & 200 & 300 & 400 & 500
\\
\hline \hline \lambda_{t^{\prime}}^{(I)}\times 10^{-2}
&-8.81&-7.03&-6.27&-5.85&-5.41&-5.17&-5.03 \\ \hline
\lambda_{t^{\prime}}^{(II)}\times 10^{-3}
&-7.76&-6.18&-5.51&-5.13&-4.74&-4.53&-4.41
\\ \hline
\multicolumn{8}{|c|}{BR(\bsGAM)=3.15\times 10^{-4}}\\
\hline {m_{t^\prime}~(GeV)} & 75 & 100 & 150 & 200 & 300 & 400 & 500
\\
\hline \hline \lambda_{t^{\prime}}^{(III)}\times 10^{-2}
&-9.29&-7.41&-6.61&-5.70&-5.45&-5.30&-5.09 \\ \hline
\lambda_{t^{\prime}}^{(IV)}\times 10^{-3}
&-3.03&-2.41&-2.14&-1.99&-1.84&-1.76&-1.71
\\ \hline
\multicolumn{8}{|c|}{BR(\bsGAM)=3.37\times 10^{-4}}\\
\hline {m_{t^\prime}~(GeV)} & 75 & 100 & 150 & 200 & 300 & 400 & 500
\\
\hline \hline \lambda_{t^{\prime}}^{(V)}\times 10^{-2}
&-9.49&-7.56&-6.74&-6.29&-5.81&-5.56&-5.41 \\ \hline
\lambda_{t^{\prime}}^{(VI)}\times 10^{-3}
&-1.07&-0.84&-0.75&-0.69&-0.64&-0.61&-0.59
\\ \hline\hline
\end{array}
$$
\caption{The numerical values of $\lambda_{t^{\prime}}$ for different values of the $m_{t^\prime}$ --quark mass and
experimental values . The
superscripts $(I),...,(VI)$ correspond to first and last
solutions of Eq. (9) with the approximation of Eq. (6). Notice that in this table real values of
 $\lambda_{t^{\prime}}$ is presented only.
In table 3 imaginary parts can be found.}
\renewcommand{\arraystretch}{1}
\addtolength{\arraycolsep}{-5pt}
\end{table}

\begin{table}[ht]
\renewcommand{\arraystretch}{2}
\addtolength{\arraycolsep}{5pt}
$$
\begin{array}{|c|c|c|c|c|c|c|c|}\hline
\multicolumn{8}{|c|}{BR(\bsGAM)=2.66\times 10^{-4}}\\
\hline {m_{t^\prime}~(GeV)} & 75 & 100 & 150 & 200 & 300 & 400 & 500
\\
\hline \hline \lambda_{t^{\prime}}^{(I)}\times 10^{-2}
&0.28&0.17&0.13&0.11&0.09&0.08&0.07 \\ \hline
\lambda_{t^{\prime}}^{(II)}\times 10^{-3}
&-0.17&-0.14&-0.13&-0.12&-0.11&-0.11&-0.10
\\ \hline
\multicolumn{8}{|c|}{BR(\bsGAM)=3.15\times 10^{-4}}\\
\hline {m_{t^\prime}~(GeV)} & 75 & 100 & 150 & 200 & 300 & 400 & 500
\\
\hline \hline \lambda_{t^{\prime}}^{(III)}\times 10^{-2}
&-0.31&-0.19&-0.15&-0.11&-0.09&-0.08&-0.07 \\ \hline
\lambda_{t^{\prime}}^{(IV)}\times 10^{-3}
&-2.10&-1.68&-1.50&-1.41&-1.30&-1.25&-1.21
\\ \hline
\multicolumn{8}{|c|}{BR(\bsGAM)=3.37\times 10^{-4}}\\
\hline {m_{t^\prime}~(GeV)} & 75 & 100 & 150 & 200 & 300 & 400 & 500
\\
\hline \hline \lambda_{t^{\prime}}^{(V)}\times 10^{-2}
&-0.32&-0.19&-0.15&-0.13&-0.11&-0.09&-0.09 \\ \hline
\lambda_{t^{\prime}}^{(VI)}\times 10^{-3}
&-2.1&-1.6&-1.5&-1.4&-1.3&-1.24&-1.21
\\ \hline\hline
\end{array}
$$
\caption{Imaginary parts of  $\lambda_{t^{\prime}}$  values, presented in table 2. }
\renewcommand{\arraystretch}{1}
\addtolength{\arraycolsep}{-5pt}
\end{table}

In order to check the  consistency of the results of present work one can demand
$\lambda_{t^\prime}$ values to satisfy the unitarity condition.
If we impose the unitarity condition of the CKM matrix  we then
have
 \bea \lambda_{u}+ \lambda_{c}+ \lambda_{t}+ \lambda_{t^\prime}=0 ~.
\label{unity}\eea
With the values of the CKM matrix elements in
the SM \cite{Ref29}, the sum of the first three terms in Eq. (\ref{unity})
is about $7.6 \times 10^{-2}$, where the error in sum of first
three terms  is about $\pm 0.6 \times 10^{-2}$. By substituting the
values of $\lambda_{t^\prime}$ from tables 1 and 2,
we observe that the sum of the four terms on the left--hand side
of Eq. (\ref{unity}) may get  very close to zero or diverge from the prediction of SM .
When $\lambda_{t^\prime}$ is very close to the sum of
the first three terms, but with opposite sign, this is a very desirable result.
Using table 2 for $m_{t^\prime}=100 $ GeV and
 the experimental branching ratio $3.37 \times 10^-4$, our prediction reads $\lambda^{(V)}_{t^\prime}=-7.56\times 10^{-2}$.
On the other hand the same prediction contains  an imaginary part
$-0.19\,i \times 10^{-2}$,
 which may  be absorbed within the error range.
 In other words, results presented
 in table (2)  satisfy unitarity constrain to a good extend. Nevertheless,
  it is a matter of taste to accept or reject $\lambda_{t^\prime}$ values, according to unitarity condition.
  Because,
 it is possible that , existence of extra generations can  affect present constraints on $V_{CKM}$ to a certain extend,
 and hence, constraints may get relaxed \cite{hamza}, which is beyond the scope
 of this work. From this respect it is hard to  claim that all results presented here can satisfy unitarity.
Nevertheless, in order to give the full picture, we did not
exclude  regions  that violates unitarity.

\subsection{Differences in  the definitions of $\lambda_{t^\prime}$}

In order to explain  the difference,
 on the results of the two different  approaches given in Eq. (6)  and Eq. (12) or tables (1) and (2),
we can perform the analysis in LO, to extract the value of the fourth generation
 CKM matrix element $\lambda_{t^\prime}$.
 Following \cite{Ref19}, one can  use the
experimental results of the decays $BR(\bsGAM)$
and $BR(B \to X_c e \bar{v}_e)$, as in \cite{Ref27}. In order to
reduce the uncertainties arising from $b$ quark mass,
 consider the following ratio

\begin{equation}
R = \frac{Br(B \to X_s \gamma)}{Br(B \to X_c e \bar{v}_e)} \,.
\end{equation}
In leading logarithmic  approximation, for low energy scale
approximation ratio can be written as

\begin{equation}\label{r}
R = \alpha_{m}\vert C^{\rm eff}_7(\mu_b) \vert^2
\end{equation}

where $\alpha_{m}=\frac{ \vert V^*_{ts} V_{tb} \vert^2}{\vert V_{cb} \vert^2} \frac{6 \alpha}{\pi f(\hat{m}_c) \kappa(\hat{m}_c)}$
 , the phase factor $f(\hat{m}_c)$ and ${\cal O}(\alpha_s)$,
 QCD correction factor $\kappa(\hat{m}_c)$  of $b \to c l \bar{\nu}$ are given in ref.\cite{Buras2}.
 Using the LO definition of $C^{eff}_7(\mu_b)$ one can write \cite{burasun},
\bea
C^{eff}_7(\mu_b)=\eta^{16/23} C^{eff}_7(\mu_W)+\frac{8}{3} (\eta^{14/23}-\eta^{16/23}) C^{eff}_8(\mu_W)+C^{eff}_2(\mu_W) \sum^{8}_{i=1}h_i \eta^{a_i}
\eea
for the present purpose, which can be written as
\bea\label{purpose}
C^{eff}_7(\mu_b)=\eta_{1}\, C^{eff}_7(\mu_W)+\eta_{2}\, C^{eff}_8(\mu_W)+ \eta_{3}\,C^{eff}_2(\mu_W)
\eea
When the effect of 4-generation it is defined as Eq. (12)
\bea
 C_{7,8}^{eff}(\mu_b) &=& C_{7,8}^{SM}(\mu_b) + \frac{\lambda_{t^\prime} }
{\lambda_{t}}
 C_{7,8}^{New} (\mu_b) ~, \nnb \\
   \eea
solution of  Eq. (\ref{r}) for $\lambda_{t^\prime}$  can be written as follows
\begin{equation}\label{sol}
\lambda_{t^\prime}^\pm = \left[ \pm \sqrt{ \frac{R}{\alpha_{m} }} -
C_7^{\rm SM}(\mu_b) \right] \frac{\lambda_t}{C_7^{\rm New}(\mu_b)} \,.
\end{equation}
whereas in the case of the following approach ( Eq. (12))
\bea
 C_{7,8}^{eff}(\mu_W) &=& C_{7,8}^{SM}(\mu_W) + \frac{\lambda_{t^\prime}}
{\lambda_{t}} C_{7,8}^{New} (\mu_W) ~, \nnb \\
\eea
Eq. (\ref{sol}) is modified into  the following form
\begin{equation}
\label{solu2} \lambda_{t^\prime}^\pm = \left[ \pm \sqrt{
\frac{R}{\alpha_{m}}} - C_7^{\rm SM}(\mu_b) \right]
\frac{\lambda_t}{\left[\eta_{1}\,C_7^{\rm New}(\mu_b)+
\eta_{2}\,C_8^{\rm New}(\mu_b)\right]} \,.
\end{equation}
This analysis can also be performed for NLO expressions. By
comparing Eq. (\ref{sol})  and Eq. (\ref{solu2}) the difference
in tables (1) and and (2) can be inferred. It should be stressed
that, for Eq.(17), possibility of a complex solution for
$\lambda_{t^\prime}$ should not be excluded.

\subsection{Direct CP violation in $\bsGAM$}

Observation of   CP violation  in $\bsGAM$ is  attractive, because  it could lead to
an evidence related with the new physics. Theoretical predictions
for $\bsGAM$ can be written as
\begin{equation}
\label{SMpredict} \nonumber A_{CP}({B \rightarrow  X_{s} \,
\gamma}) = \frac{\Gamma(\bar B \rightarrow X_{s}\gamma)
     -\Gamma(B \rightarrow  X_{\bar s}\gamma)}
     {\Gamma(\bar B \rightarrow  X_{s} \gamma)
     +\Gamma(B \rightarrow  X_{\bar s}\gamma)}\,.
\end{equation}
 Numerically,
prediction of the SM is \cite{tobiash}
\begin{equation}
  A_{CP}({B \rightarrow  X_s \gamma}) \approx 0.6 \%, \qquad
\label{SMnumber}
\end{equation}
when the best-fit values for the CKM parameters \cite{CKMfit} are
used. From the experimental side, we have  the CLEO  measurement of the CP
asymmetry in the $b \rightarrow  s \gamma$ decays \cite{cleoCP},
 \beq
A_{CP}(B \rightarrow  X_s \gamma) = (-0.079 \pm 0.108 \pm
0.022) \times (1.0 \pm 0.030) \, ,
\eeq

We used the CP asymmetry formulae  to look for 4 generation effects  \cite{tobiash},

\begin{eqnarray}
\label{SMCPexplicit} A_{CP}({B \rightarrow  X_{s} \, \gamma})
\simeq \frac{10^{-2}}{|C_7|^2} ( 1.17 \times {\rm Im}\left[C_2C_7^*\right]
                - 9.51 \times {\rm Im}\left[C_8 C_7^*\right]   \\
+0.12\times {\rm Im}\left[C_2C_8^*\right] -9.40 \times {\rm Im}\left[\epsilon_{s}C_2\left(
                C_7^*-0.013 \; C_8^*\right)\right] ) \nonumber;\\
\epsilon_s = \frac{V_{us}^*V_{ub}}{V_{ts}^*V_{tb}} \simeq
-\lambda^2(\rho-i\eta).
 \quad \quad
\nonumber
\end{eqnarray}

As it is stated in the same reference, the large coefficient of the second term in (\ref{SMCPexplicit})
is very attractive.
\begin{figure}
\vskip 1.5 cm \includegraphics{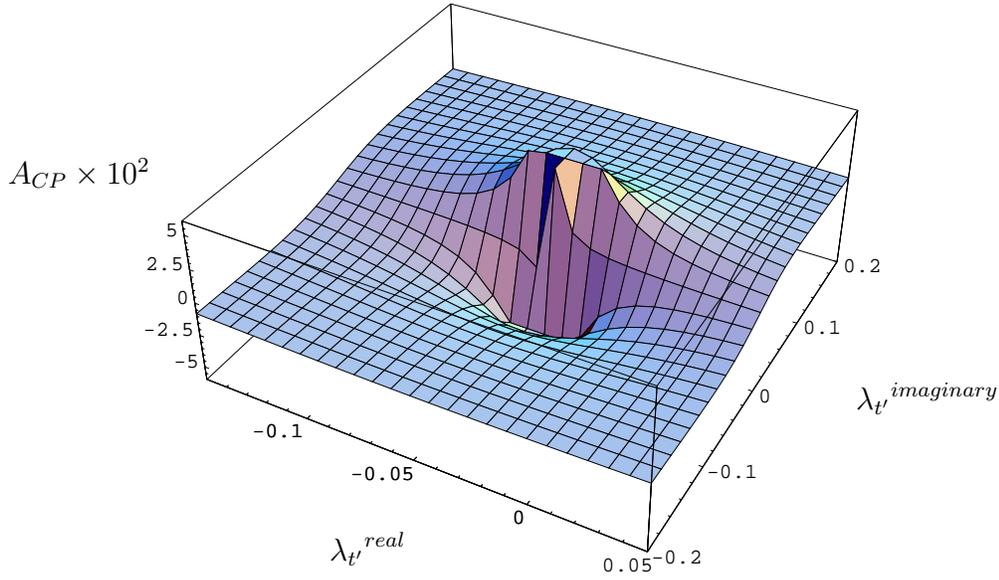} \vskip 0 cm \hskip 0 cm
\title{$A_{CP}\times 10^2$}
 \vskip 2.5 cm \hskip +11 cm
\title{~~${\lambda_{t^{\prime}}}^{imaginary}$}
\vskip 1.5 cm \hskip 4 cm
\title{~~${\lambda_{t^{\prime}}}^{real}$}
   \vskip .5 cm \caption{$A_{CP}(B\rightarrow X_s\gamma)$ for $m_{t^{\prime}}=~75$.}
\end{figure}
We observed that, enhanced chromomagnetic dipole contribution, $C_8$, induces a
large direct CP violation in the decay $B \rightarrow X_s
\gamma$. This is due to complex phases of $\lambda_{t^\prime}$, which in result affects $C_7,~C_8$. Such an enhancement of the chromomagnetic
contribution may  lead to a natural explanation of the
phenomenology of semileptonic $B$ decays  and  charm production
in $B$ decays \cite{KaganNeubert,tobiash}.

\begin{figure}
\vskip 1.5 cm \includegraphics{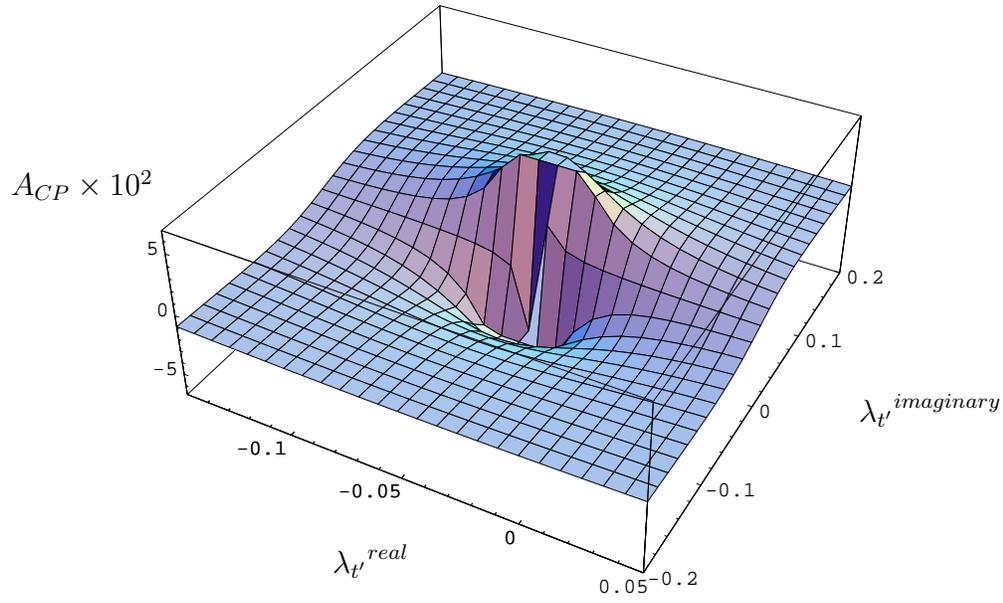}
\vskip 0 cm \hskip 0 cm
\title{$A_{CP}\times 10^2$} \vskip 2.5 cm \hskip +11 cm
\title{~~${\lambda_{t^{\prime}}}^{imaginary}$}
\vskip 1.5 cm \hskip 4 cm
\title{~~${\lambda_{t^{\prime}}}^{real}$}
   \vskip .5 cm \caption{$A_{CP}(B\rightarrow X_s\gamma)$ for $m_{t^{\prime}}=~50$.}
\end{figure}

Notice that in $A_{CP}$ figures, when  the real values of $\lambda_{t^\prime}$  is
 around $-6 \times 10^{-2}$, even for very small imaginary parts, peak values of $A_{CP}$ can be  observed.
  Evolution  of $A_{CP}({B \rightarrow  X_{s} \, \gamma})$ is
  presented in figures $\{7,8,9,10\}$. CP asymmetry is not
  sensitive to very heavy $m_{t^\prime}$ quark masses.

\begin{figure}
\vskip 1.5 cm \includegraphics{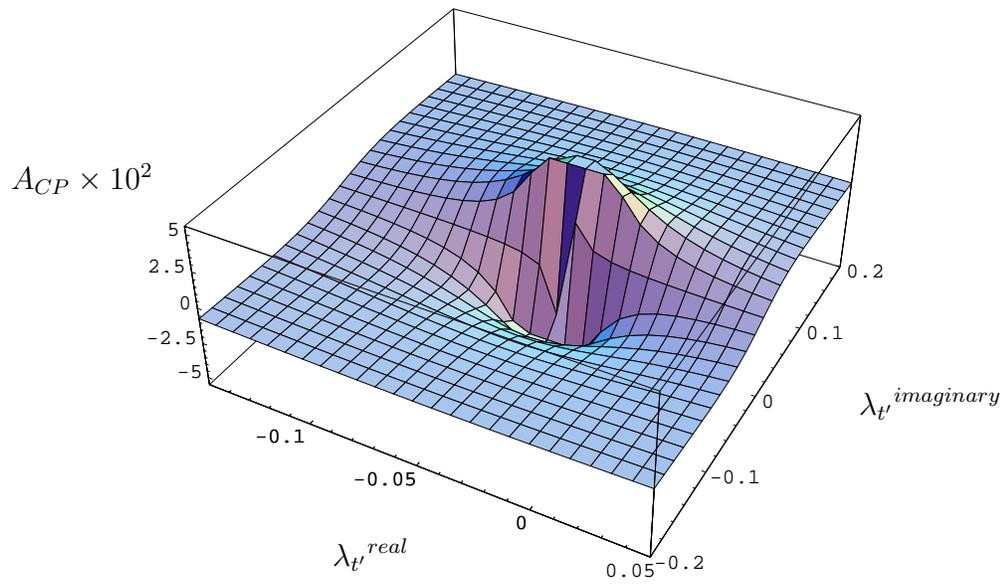}
\vskip 0 cm \hskip 0 cm
\title{$A_{CP}\times 10^2$}
 \vskip 2.5 cm \hskip +11 cm
\title{~~${\lambda_{t^{\prime}}}^{imaginary}$}
\vskip 1.5 cm \hskip 4 cm
\title{~~${\lambda_{t^{\prime}}}^{real}$}
   \vskip .5 cm \caption{$A_{CP}(B\rightarrow X_s\gamma)$ for $m_{t^{\prime}}=~100$.}
\end{figure}

\begin{figure}
\vskip 1.5 cm \includegraphics{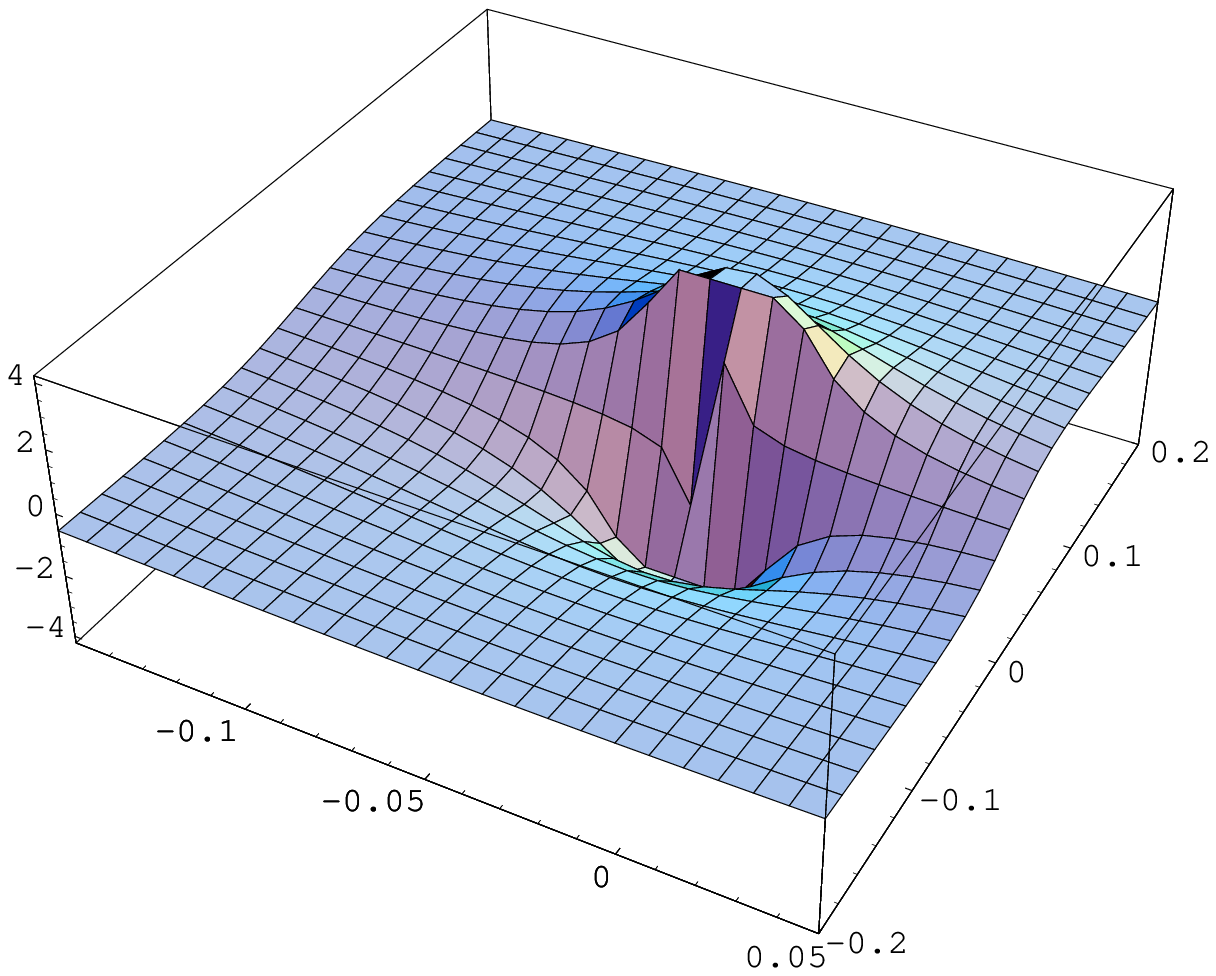}
\vskip 0 cm \hskip 0 cm
\title{$A_{CP}\times 10^2$}
 \vskip 2.5 cm \hskip +11 cm
\title{~~${\lambda_{t^{\prime}}}^{imaginary}$}
\vskip 1.5 cm \hskip 4 cm
\title{~~${\lambda_{t^{\prime}}}^{real}$}
   \vskip .5 cm \caption{$A_{CP}(B\rightarrow X_s\gamma)$ for $m_{t^{\prime}}=~300$.}
\end{figure}

\begin{figure}
\vskip 1.5 cm \includegraphics{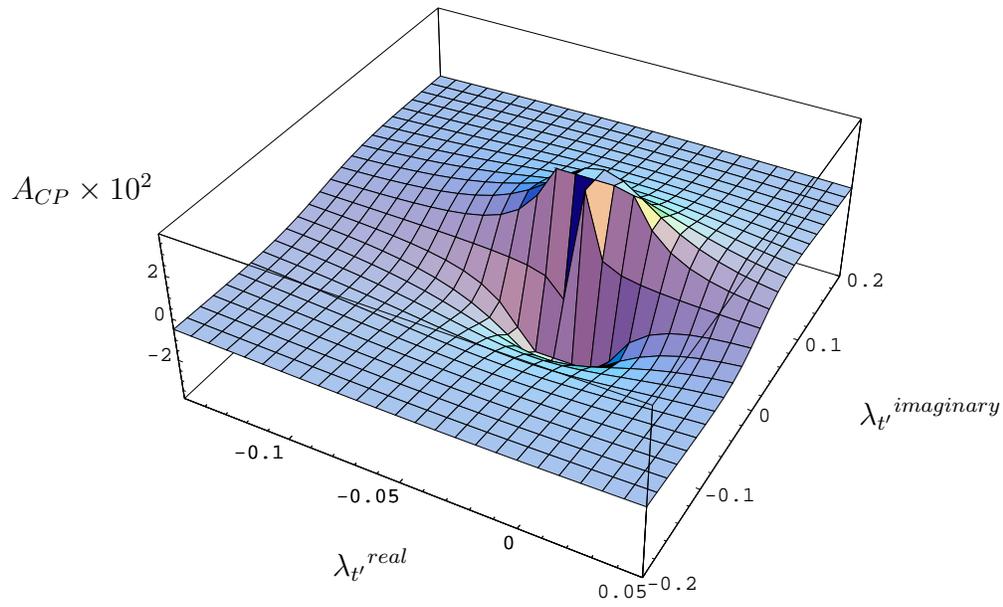}
\vskip 0 cm \hskip 0 cm
\title{$A_{CP}\times 10^2$}
 \vskip 2.5 cm \hskip +11 cm
\title{~~${\lambda_{t^{\prime}}}^{imaginary}$}
\vskip 1.5 cm \hskip 4 cm
\title{~~${\lambda_{t^{\prime}}}^{real}$}
   \vskip .5 cm \caption{$A_{CP}(B\rightarrow X_s\gamma)$ for $m_{t^{\prime}}=~500$.}
\end{figure}

\section{Conclusion}

To summarize, the $\bsGAM$ decay has a clean experimental and
theoretical base, very sensitive to the various extensions of the
Standard Model, can be used to constrain the fourth generation
model. In the present work, this decay is studied in the SM with
the four generation model. The solutions of the fourth generation
CKM factor $\lambda_{t^\prime}$ have been
obtained. It is observed   that different  choices of the factor
$\lambda_{t^\prime}$, could
be very informative, especially due to new CP violation effects,
in searching new physics.

CP asymmetry in the $\bsGAM$ decay can be enhanced up to $5 ~\%$,
which is ten times larger compared to the SM prediction. Hence it
could be mentioned among the probes of new physics, especially in
the case of fourth generation.


\end{document}